\documentstyle[11pt]{article}
\begin{document}
\renewcommand{\refname}{References.}
\newcommand{\Nf}{N_{\!f}}
\newcommand{\partialslash}{\partial \!\!\! /}
\newcommand{\kslash}{k \!\!\! /}
\newcommand{\half}{{\mbox{\small{$\frac{1}{2}$}}}}
\title{Electron mass anomalous dimension at $O(1/\Nf^2)$ in quantum
electrodynamics.}
\author{J.A. Gracey, \\ Department of Applied Mathematics and Theoretical
Physics, \\ University of Liverpool, \\ P.O. Box 147, \\ Liverpool, \\
L69 3BX, \\ United Kingdom.}
\date{}
\maketitle
\vspace{5cm}
\noindent
{\bf Abstract.} The critical exponent corresponding to the renormalization of
the composite operator $\bar{\psi}\psi$ is computed in quantum electrodynamics
at $O(1/\Nf^2)$ in arbitrary dimensions and covariant gauge at the non-trivial
zero of the $\beta$-function in the large $\Nf$ expansion and the exponent
corresponding to the anomalous dimension of the electron mass which is a gauge
independent object is deduced. Expanding in powers of $\epsilon$ $=$ $2$ $-$
$d/2$ we check it is consistent with the known three loop perturbative
structure and determine the subsequent coefficients in the coupling constant
expansion to {\em all} orders in $\overline{\mbox{MS}}$.

\vspace{-18cm}
\hspace{10cm}
{\bf LTH-316}
\newpage
The fundamental functions of the renormalization group equation, such as the
$\beta$-function and mass anomalous dimension, are crucial ingredients in
determining the properties of the Green's functions of a renormalizable
quantum field theory and how they depend on the energy scale. They are
ordinarily determined by renormalizing Green's functions at successive orders
in perturbation theory where the coupling constant, $g$, is assumed to be
small. Invariably one finds that calculations become exceedingly difficult at a
certain order, due to the large number of Feynman graphs to be evaluated, which
is therefore tedious, and the intricate nature of some of the integrals which
can arise. Thus the renormalization group functions are only known to the first
few orders for most theories. For example, the $\beta$-function of QED has
recently been calculated to fourth order in $\overline{\mbox{MS}}$ in \cite{1}
by use of computer algebra packages and the wave function and mass
renormalization functions are known to third order, by restricting the QCD
results of \cite{2} to the abelian case, which extended the earlier two loop
calculations of \cite{3,4}. Clearly it is important to have methods which
circumvent these calculational difficulties and allows one to gain an insight
into the perturbative structure of these functions to very high orders in the
coupling constant.

One way to achieve this is by computing the functions in terms of a different
expansion parameter. For example, in theories which possess $N$ fundamental
fields, such as $\Nf$ electrons in QED, one can use $1/N$ as an alternative
expansion parameter. In the earlier work of \cite{5} and \cite{6} the
$\beta$-function and mass renormalization, which are gauge independent
quantities \cite{7,8}, were computed at leading order in large $\Nf$ in QED.
This was achieved by explicitly computing the simple poles of the leading order
Green's functions, in dimensional regularization, where the photon propagator
was replaced by a chain of one loop bubble graphs. The simple poles were then
absorbed minimally into a renormalization constant and the
$(4-2\epsilon)$-dimensional perturbation series of $\beta(g)$ and $\gamma_m(g)$
were deduced, which gave the $O(1/\Nf)$ coefficients of each function to all
orders in $g$. Whilst this was successful in giving agreement with what had
previously been calculated in $\overline{\mbox{MS}}$ and also provided new
coefficients, it is certainly not easy to extend that analysis to probe the
perturbation series more deeply.

An alternative approach is to use the large $\Nf$ self consistency method
introduced in \cite{9,10} where $d$-dimensional critical exponents are computed
at the non-trivial zero of the $\beta$-function. As the theory is finite and
possesses a conformal symmetry there, the renormalization group equation
simplifies to the extent that the critical exponents one computes are related
to the critical renormalization group functions which also allows one to
extract all orders information. The beauty of applying the method to a four
dimensional gauge theory such as we do here for QED is that by using dressed
propagators one substantially reduces the number of Feynman diagrams to be
analysed and crucially it is possible to achieve results at $O(1/\Nf^2)$.
Previously using these techniques in QED the electron wave function exponent
$\eta$ was deduced at $O(1/\Nf)$ in the Landau gauge in \cite{11} and later at
$O(1/\Nf^2)$ in \cite{12}. The original $\beta$-function calculation of
\cite{6} was reproduced more concisely in \cite{13} at $O(1/\Nf)$ and the gauge
independence was checked in \cite{14} by computing in an arbitrary covariant
gauge. The significance of the agreement of \cite{13} with \cite{6} is that it
establishes that the exponents calculated by the self consistency method of
\cite{9,10} encode $\overline{\mbox{MS}}$ information.

In this letter, we extend the work of \cite{6} by computing the electron mass
anomalous dimension, $\gamma_m(g)$, to $O(1/\Nf^2)$ by determining the relevant
critical exponent. Its calculation is necessary as an independent check on the
three loop result of \cite{2} as well as providing the remainder of the
perturbation series for $\gamma_m(g)$ at $O(1/\Nf^2)$. This is required for
future four loop QED and QCD calculations and will also serve as the foundation
for finding the $O(1/\Nf^2)$ quark mass anomalous dimensions. Simultaneously we
deduce a result for three dimensional QED which is important since it will
provide improved estimates for a gauge independent quantity which can be
measured on the lattice. This will give a useful check for the Monte Carlo
algorithms used to simulate problems such as the existence or otherwise of a
chirally symmetric phase. Finally, we draw attention to the fact that the
procedure we follow here is based very much on the $O(1/N^2)$ calculation of
the mass operator in the $O(N)$ Gross Neveu model, \cite{15}.

To compute critical exponents in large $\Nf$ in QED we use the lagrangian in
the form, \cite{11},
\begin{equation}
L ~=~ i \bar{\psi}^i \partialslash \psi^i + A_\mu \bar{\psi}^i \gamma^\mu
\psi^i - \frac{F^2_{\mu\nu}}{4e^2} - \frac{(\partial_\mu A^\mu)^2}{2\xi e^2}
\end{equation}
where $\psi^i$ is the electron field, $1$ $\leq$ $i$ $\leq$ $\Nf$, $F_{\mu\nu}$
$=$ $\partial_\mu A_\nu$ $-$ $\partial_\nu A_\mu$, $\xi$ is the covariant
gauge parameter and $A_\mu$ is the $U(1)$ gauge field. The electron coupling
constant, $e$, has been rescaled into the gauge field kinetic term to ensure
that the vertex has the correct structure for applying the method of
uniqueness of \cite{16} which is the main technique for computing the massless
Feynman integrals which arise. The location of the critical coupling, $g_c$,
in whose neighbourhood we will solve (1) is defined as the non-trivial zero of
the $d$-dimensional $\beta$-function. It has been calculated in
$\overline{\mbox{MS}}$ to fourth order in \cite{1} in
$(4-2\epsilon)$-dimensions using dimensional regularization as
\begin{eqnarray}
\beta(g) &=& (d-4)g + \frac{2\Nf}{3} g^2 + \frac{\Nf}{2} g^3
- \frac{\Nf(22\Nf+9)}{144} g^4 \nonumber \\
&-& \frac{\Nf}{64} \left[ \frac{616\Nf^2}{243} + \left( \frac{416\zeta(3)}{9}
- \frac{380}{27} \right) \Nf + 23 \right]g^5 + O(g^6)
\end{eqnarray}
where we use the conventions of \cite{17} but have defined $g$ $=$
$(e/2\pi)^2$. Thus to deduce information on the $(4-2\epsilon)$-dimensional
perturbation series from the exponent we will calculate, (2) implies that then
\begin{equation}
g_c ~ \sim ~ \frac{3\epsilon}{\Nf} - \frac{27\epsilon^2}{4\Nf^2}
+ \frac{99\epsilon^3}{16\Nf^2} + \frac{77\epsilon^4}{16\Nf^2}
\end{equation}
In the critical region the field theory has a conformal symmetry and so the
propagators and Green's functions obey a simple power law behaviour. For
instance, in momentum space, as $k^2$ $\rightarrow$ $\infty$, the propagators
of (1) satisfy, \cite{11},
\begin{equation}
\psi(k) ~ \sim ~ \frac{\tilde{A}\kslash}{(k^2)^{\mu-\alpha}} ~~,~~
A_{\nu\sigma}(k) ~ \sim ~ \frac{\tilde{B}}{(k^2)^{\mu-\beta}} \left[
\eta_{\nu\sigma} - (1-\xi) \frac{k_\nu k_\sigma}{k^2} \right]
\end{equation}
where $\tilde{A}$ and $\tilde{B}$ are the respective $k$-independent amplitudes
and $\alpha$ and $\beta$ are the critical exponents of each field and their
canonical and anomalous dimensions are defined to be, \cite{11},
\begin{equation}
\alpha ~=~ \mu - 1 + \half \eta ~~,~~ \beta ~=~ 1 - \eta - \chi
\end{equation}
where $\eta$ is the electron anomalous dimension and $\chi$ is the anomalous
dimension of the $3$-vertex. Each is $O(1/\Nf)$ and depends on $d$ $=$ $2\mu$.
The former corresponds through an examination of the critical point RGE to the
electron wave function renormalization which is known to third order in
\cite{2} and $\eta$ has been deduced at $O(1/\Nf^2)$ in the Landau gauge in
\cite{12}. Although it is a gauge dependent quantity it must always be
calculated first within the self consistency formalism of \cite{9} as it plays
a central role in the computation of other exponents.

To compute the electron mass exponent, $\gamma_m(g_c)$, we follow the same
strategy as the two and three loop perturbative approaches of [2-4]. There
$\gamma_m(g)$ was calculated indirectly by first renormalizing the composite
operator $\bar{\psi}\psi$ and then using
\begin{equation}
\gamma_m(g) ~=~ \gamma_{\bar{\psi}\psi}(g) + \gamma_2(g,\xi)
\end{equation}
where $\gamma_2(g,\xi)$ is the wave function renormalization constant which
differs in overall sign to the definition given in \cite{2}. In exponent
language (6) translates into
\begin{equation}
\gamma_m(g_c) ~=~ \eta + \gamma_{\bar{\psi}\psi}(g_c)
\end{equation}
Since both $\eta$ and $\gamma_{\bar{\psi}\psi}(g_c)$ are gauge dependent
the cancellation of the parameter $\xi$ when we calculate with an arbitrary
covariant gauge will provide a very stringent check on our computation before
checking with \cite{2}. For completeness, in the same notation as (2),
\cite{2},
\begin{eqnarray}
\gamma_m(g) &=& - \, \frac{3}{2}g + \frac{(20\Nf-9)}{48}g^2 \nonumber \\
&+& \left[ \frac{140\Nf^2}{27} + (46 - 48\zeta(3))\Nf - \frac{129}{2}
\right] \frac{g^3}{32} + O(g^4)
\end{eqnarray}

We will now detail the leading order analysis to reproduce the result of
\cite{6} but in the critical point self consistency approach which will also
illustrate the simplicity of the method and will allow us to elaborate on
several technical points in preparation for the $O(1/\Nf^2)$ calculation.
First, we have to introduce a regularization to handle the infinities which
will arise in the calculation. This is achieved by shifting the exponent
$\chi$ by $\chi$ $\rightarrow$ $\chi$ $+$ $\Delta$ where $\Delta$ is an
infinitesimal quantity, \cite{12}. Once this is introduced the Green's
functions which one calculates in $d$-dimensions will have a three part
structure, \cite{18,19}. It will involve terms which have poles in $\Delta$
which are removed by a conventional renormalization, \cite{18}. This leaves
$O(1)$ terms which are multiplied by $\ln p^2$, where $p$ is the external
momentum, and $O(\Delta)$ terms which vanish as the regularization is lifted.
As we are working at a critical point where $p^2$ $\rightarrow$ $\infty$ the
presence of $\ln p^2$ terms will spoil the scaling behaviour. General
arguments, however, show that their sum will be of the form
$(p^2)^{\half \gamma_{\bar{\psi}\psi}(g_c)}$, \cite{18,19}. Therefore,
expanding this in powers of $1/\Nf$ and choosing $\gamma_{\bar{\psi}\psi}
(g_c)$ appropriately at each order will remove these $\ln p^2$ terms to
leave a Green's functions with sensible scaling behaviour. Indeed this
observation which was discussed originally in \cite{18} was also used to
renormalize the $2$-point function of QED in \cite{12,20}. In other words to
deduce $\gamma_{\bar{\psi}\psi}(g_c)$ at each order in $1/\Nf$ one computes the
Feynman diagrams with dressed propagators which contribute to the inclusion of
$\bar{\psi}\psi$ as a composite operator in some Green's function, which for
this letter will be $\langle \psi [\bar{\psi}\psi] \bar{\psi} \rangle$,
isolates the $\ln p^2$ terms after $\Delta$-renormalization and sums their
coefficients.

At leading order there is only one Feynman graph to consider which is given in
fig. 1. The circle with a cross indicates a $\bar{\psi}\psi$ insertion.
Substituting the asymptotic scaling forms of the propagators of (4) for the
lines of fig. 1 and computing the trivial one loop integral, it is
\begin{equation}
z(2\mu-1+\xi) a(\mu-1+\Delta) a(1) a(\mu-\Delta) (p^2)^{- \, \Delta}
\end{equation}
where we have defined $a(\alpha)$ $=$ $\Gamma(\mu-\alpha)/\Gamma(\alpha)$ for
all $\alpha$ and set $z$ $=$ $\tilde{A}^2\tilde{B}$. Expanding (9) in powers of
$\Delta$ one obtains
\begin{equation}
\frac{(2\mu-1+\xi)z_1}{\Delta \Gamma(\mu) \Nf} \left[ 1 - \Delta \ln p^2
+ \frac{\Delta}{(\mu-1)} + O(\Delta^2) \right]
\end{equation}
where $z$ $=$ $\sum_{i=1}^\infty z_i /\Nf^i$ and $z_1$ is deduced from the
leading order computation of $\eta$ in \cite{11} as
\begin{equation}
z_1 ~=~ \frac{\Gamma(2\mu)}{8\Gamma^2(\mu) \Gamma(2-\mu)}
\end{equation}
Thus,
\begin{equation}
\gamma_{\bar{\psi}\psi , 1}(g_c) ~=~ - \, \frac{\mu(2\mu-1+\xi)
\eta^{\mbox{o}}_1}{(\mu-2)(2\mu-1)}
\end{equation}
where
\begin{equation}
\eta^{\mbox{o}}_1 ~=~ \frac{(2\mu-1)(\mu-2)\Gamma(2\mu)}{4\Gamma^2(\mu)
\Gamma(\mu+1)\Gamma(2-\mu)}
\end{equation}
which corresponds to the Landau gauge expression for $\eta_1$. It is a trivial
exercise to repeat the computation of $\eta_1$ for non-zero $\xi$ to deduce
\begin{equation}
\eta^\xi_1 ~=~ \frac{[(2\mu-1)(\mu-2)+\xi\mu]\eta^{\mbox{o}}_1}{(2\mu-1)
(\mu-2)}
\end{equation}
which from (7) implies
\begin{equation}
\gamma_{m , 1}(g_c) ~=~ - \, \frac{2 \eta^{\mbox{o}}_1}{(\mu-2)}
\end{equation}
in agreement with \cite{6}, but deduced by considering only one Feynman graph.

The way to proceed to $O(1/\Nf^2)$ is now clear and involves first of all
computing the contributions from fig. 1 since its lines are represented by
propagators which possess $\Nf$ dependent exponents as well as the obvious
counterterm graphs. However, the main effort lies in considering the higher
order graphs of fig. 2. The fourth graph can be determined by a simple
extension of the one loop integral of fig. 1 and it contributes
\begin{equation}
- \, \frac{3\mu^2(2\mu-1+\xi)^2 (\eta_1^{\mbox{o}})^2}{2(\mu-1)(\mu-2)^2
(2\mu-1)^2}
\end{equation}
to $\gamma_{\bar{\psi}\psi , 2}(g_c)$. To calculate the remaining graphs one
first maps each integral into coordinate space through the
Fourier transform
\begin{equation}
\frac{1}{(x^2)^\alpha} ~=~ \frac{a(\alpha)}{2^{2\alpha}\pi^\mu} \int_k
\frac{e^{ikx}}{(k^2)^{\mu-\alpha}}
\end{equation}
Since the third graph of fig. 2 then corresponds to an insertion on the
completely internal electron line of the second order correction to the
electron self energy it is $\Delta$-finite in coordinate space, \cite{20}.
Also, the values of the first two graphs are equivalent and after mapping to
coordinate space they correspond to a self energy graph where the insertion is
now on an electron line joining to an external vertex. It is not instructive to
reproduce the tedious algebra connected with the calculation of these graphs
aside from remarking that the techniques of integration by parts, \cite{14},
subtractions, \cite{10}, and uniqueness, \cite{16}, we used have already been
discussed in the computation of the electron self energy at second order,
\cite{12,20}. After a substantial amount of algebra we found that the first
graph of fig. 2 contributes
\begin{eqnarray}
&-& \frac{\mu(\eta_1^{\mbox{o}})^2}{(2\mu-1)^2(\mu-2)^2} \left[
\frac{(2\mu-1+\xi)[(2\mu-1)(\mu-2)+\xi\mu]}{(\mu-1)} \right. \nonumber \\
&& \quad \quad \quad \quad \quad \quad \quad \quad - \, \left.
\frac{(2\mu-1)(\mu-1)(1-\xi)}{\mu} \right]
\end{eqnarray}
to $\gamma_{\bar{\psi}\psi , 2}(g_c)$, whilst the third gave
\begin{equation}
- \, \frac{\mu^2(\eta_1^{\mbox{o}})^2}{2(\mu-1)(2\mu-1)^2(\mu-2)^2}
[(2\mu-1+\xi)(2\mu-3-\xi) - 4\xi(\mu-1)]
\end{equation}
The contributions from the graph of fig. 1 which involves the various
counterterm pieces are
\begin{eqnarray}
&& \frac{\mu(2\mu-1+\xi)(\eta_1^{\mbox{o}})^2}{(\mu-1)(2\mu-1)^2(\mu-2)^2}
\left[ \frac{}{} (2\mu-1)(3\mu-4) \right. \nonumber \\
&& \left. \quad \quad \quad \quad \quad \quad + ~ 3\xi\mu - 3\mu(\mu-1)^2
\left( \hat{\Theta} - \frac{1}{(\mu-1)^2} \right) \right]
\end{eqnarray}
where $\hat{\Theta}(\mu)$ $=$ $\psi^\prime(\mu-1)$ $-$ $\psi^\prime(1)$,
$\psi(x)$ is the logarithmic derivative of the $\Gamma$-function and we have
used $\chi_1$ $=$ $-$ $\eta_1$, \cite{14}, and
\begin{equation}
z_2 ~=~ \frac{3\mu^2(\mu-1)\Gamma(\mu) (\eta_1^{\mbox{o}})^2}
{2(\mu-2)^2(2\mu-1)^2} \left[ \hat{\Theta} - \frac{1}{(\mu-1)^2} \right]
\end{equation}
from the $x$-space calculation of $\eta_2$, \cite{12}. Thus, we have
\begin{eqnarray}
\gamma_{\bar{\psi}\psi , 2}(g_c) &=& \frac{\mu(\eta^{\mbox{o}}_1)^2}
{(2\mu-1)^2(\mu-2)^2} \left[ \frac{2(1-\xi)(2\mu-1)(\mu-1)}{\mu} 
\right. \nonumber \\
&-& \left. \!\! 2\mu(2\mu-1) - 3\mu(2\mu-1+\xi)(\mu-1)
\! \left( \! \hat{\Theta} - \frac{1}{(\mu-1)^2} \right) \right]
\end{eqnarray}

Finally, to obtain $\gamma_{m,2}(g_c)$ we need to add $\eta_2$ in an arbitrary
gauge to (21). We have deduced this by adapting the Landau gauge computation of
\cite{12} to find
\begin{eqnarray}
\eta_2^\xi &=& \frac{(\mu-1)(\eta^{\mbox{o}}_1)^2}{(\mu-2)^2(2\mu-1)^2}
\left[ \frac{2(2\mu-1)}{\mu}[(\mu-1)(\mu-3) + \xi\mu] \right. \nonumber \\
&&+~ \left. 3\mu[(2\mu-1)(\mu-2)+\xi\mu]\left( \hat{\Theta}
- \frac{1}{(\mu-1)^2} \right) \right]
\end{eqnarray}
{}From (7), (22) and (23) then we have simply
\begin{equation}
\gamma_{m,2}(g_c) ~=~ - \, \frac{6(\eta_1^{\mbox{o}})^2}{(\mu-2)^2(2\mu-1)}
\left[ \mu(\mu-1)\hat{\Theta} - \frac{\mu}{(\mu-1)} + \frac{1}{\mu}
+ \frac{4\mu}{3} - 2 \right]
\end{equation}
where it is reassuring to note that, as in (15), all $\xi$-dependence has
cancelled, which is the first check on our result.

The subsequent check is to set $\mu$ $=$ $2$ $-$ $\epsilon$ in (24) and expand
in powers of $\epsilon$ to $O(\epsilon^3)$ in order to compare with (8) at the
value of (3). We find that both (24) and (8) are in agreement and note that the
$\zeta(3)$ of (8) arises from the expansion of $\hat{\Theta}$ in (24). Thus the
gauge independent result (24) corresponds to the $O(1/\Nf^2)$ correction to
(15). Moreover, one can now deduce new coefficients which will appear in
$\gamma_m(g)$ at higher orders in $g$ by expanding $\gamma_{m,2}(g_c)$ to the
subsequent orders in $\epsilon$. The $O(1/\Nf^2)$ corrections for $g_c$ are
required for this and they are encoded in the $\beta$-function exponent
$\lambda$ $=$ $-$ $\half\beta^\prime(g_c)$ of \cite{6,14},
\begin{equation}
\lambda ~=~ \mu - 2 - \frac{(2\mu-3)(\mu-3)\eta^{\mbox{o}}_1}{\Nf}
+ O \left( \frac{1}{\Nf^2} \right)
\end{equation}
in the present notation. We record, for example, that in $\overline{\mbox{MS}}$
to fourth order in $g$, (8) will become
\begin{eqnarray}
\! \gamma_m(g) \! &=& \! - \, \frac{3g}{2} + \left[ \frac{10\Nf}{3}
- \frac{3}{2} \right] \frac{g^2}{8} + \left[ \frac{140}{27}\Nf^2
+ (46 - 48\zeta(3))\Nf - \frac{129}{2} \right] \! \frac{g^3}{32} \nonumber \\
&+& \!\! \left[ \frac{\Nf^3}{9} \! \left( \frac{83}{144} - \zeta(3) \! \right)
+ \frac{\Nf^2}{4} \! \left( \! 5\zeta(3) - 3\zeta(4) - \frac{19}{54} \! \right)
\! + \! p\Nf \! + \! q \right] \! g^4
\end{eqnarray}
where the unknown coefficients $p$ and $q$ can only be determined by an
explicit $4$-loop calculation or respectively knowledge of $O(1/\Nf^3)$ and
$O(1/\Nf^4)$ exponents. However, it is worth pointing out that one can now make
use of (26) to substantially reduce the amount of work one would have to do in
the explicit calculation. By isolating only those Feynman graphs with one and
no electron loops the residues of the simple poles in $\epsilon$ respectively
determine $p$ and $q$.

Finally, since (24) is valid in $d$-dimensions we can restrict to $\mu$ $=$
$3/2$, to obtain
\begin{equation}
\gamma_m(g_c) ~=~ - \, \frac{32}{3\pi^2\Nf} - \frac{64}{9\pi^4\Nf^2}
[3\pi^2 - 28] + O \left( \frac{1}{\Nf^3} \right)
\end{equation}
As this also is a gauge independent quantity it can be used to determine
exponents for relatively low values of $\Nf$ to compare with numerical work.

\vspace{1cm}
\noindent
{\bf Acknowledgements.} The author thanks Dr D.J. Broadhurst for suggesting
this calculation was possible, for supplying a copy of \cite{2} and for his
encouragement at various stages of the computation, as well as the organisers
of the XIIIth UK Institute for Theoretical High Energy Physics where this work
was initiated. Some of the tedious algebra was carried out using REDUCE version
3.4.

\newpage
\noindent
{\Large {\bf Figure Captions.}}
\begin{description}
\item[Fig. 1.] Leading order graph for $\gamma_{\bar{\psi}\psi}(g_c)$.
\item[Fig. 2.] Higher order graphs for $\gamma_{\bar{\psi}\psi}(g_c)$.
\end{description}
\end{document}